\documentclass[aps,prb,twocolumn,
superscriptaddress,onlinecite]{revtex4-1}

\usepackage{graphicx}
\usepackage{epstopdf}
\usepackage[scientific-notation=false]{siunitx}

\begin{document}


\title{Large magnetoresistance in the antiferromagnetic semi-metal NdSb}


\author{N. Wakeham}
\affiliation{Los Alamos National Laboratory}

\author{E. D. Bauer}
\affiliation{Los Alamos National Laboratory}

\author{M. Neupane}
\affiliation{University of Central Florida, Department of Physics}


\author{F. Ronning}
\affiliation{Los Alamos National Laboratory}



\date{\today}

\begin{abstract}
There has been considerable interest in topological semi-metals that exhibit extreme magnetoresistance (XMR). These have included materials lacking inversion symmetry such as TaAs, as well Dirac semi-metals such as Cd$_3$As$_2$. However, it was reported recently that LaSb and LaBi also exhibit XMR, even though the rock-salt structure of these materials has inversion symmetry, and the band-structure calculations do not show a Dirac dispersion in the bulk. Here, we present magnetoresistance and specific heat measurements on NdSb, which is isostructural with LaSb. NdSb has an antiferromagnetic groundstate, and in analogy with the lanthanum monopnictides, is expected to be a topologically non-trivial semi-metal. We show that NdSb has an XMR of $\sim 10^4 \%$, even within the AFM state, illustrating that XMR can occur independently of the absence of time reversal symmetry breaking in zero magnetic field. The persistence of XMR in a magnetic system offers promise of new functionality when combining topological matter with electronic correlations. We also find that in an applied magnetic field below the N$\acute{e}$el temperature there is a first order transition, consistent with evidence from previous neutron scattering work.
\end{abstract}

\maketitle
\section{Introduction}
Conventional non-magnetic metallic systems generally show a small positive magnetoresistance (MR) of a few percent, and this behavior can be well described by Boltzmann transport theory \cite{Ashcroft}. Recently, there has been a great deal of interest in non-magnetic topological metals that show an extreme magnetoresistance (XMR) of  $\sim 10^5\%$. These XMR materials are all semi-metals, with close compensation of electrons and holes, and very high carrier mobilities. However, these materials also show some unique features, because of the topology of their electronic structures. For example, Cd$_3$As$_2$ has a symmetry protected linearly dispersive band crossing and is therefore designated a Dirac semi-metal \cite{Liang2014,Neupane2014}. TaAs breaks inversion symmetry causing the doubly degenerate Dirac dispersions to split into a pair of Weyl cones \cite{Yang2015}. XMR was also recently extended to include lanthanum monopnictides \cite{Sun2016,Tafti2016,Tafti2015}. These compounds are also topologically non-trivial, although in a different sense than Weyl or Dirac semi-metals \cite{Zeng2015}. They have a simple rock salt structure with inversion symmetry, and they do not have accidental linear band crossings near the Fermi energy. Instead, they possess a band inversion, which classifies them as topological in the same vein as topological insulators, even though the bulk bands are semi-metallic as opposed to insulating. This has led to suggestions that the XMR in these materials is the result of the applied field breaking time reversal symmetry (TRS), and destroying the topological protection of the conductive states to electron back-scattering \cite{Tafti2015}.

The significance of the TRS makes it interesting to investigate similar semi-metal materials with a magnetic ground-state. In these systems, TRS will be broken even before the application of a magnetic field, and therefore one might expect significant back scattering at zero-field, in contrast to other XMR semi-metals. Also, by adding strong electronic correlations, ubiquitous in $f$-electron systems, one opens the possibility of discovering completely new states of matter with unprecedented functionality not possible in non-correlated matter \cite{Dzero2010, Hohenadler2013, Tang2011}. The interplay of magnetism and the XMR may be a route to new technological functionality in these materials. Here, we report on the resistivity and specific heat of a rare-earth monopnictide, NdSb, which is expected to be a topological semi-metal, but is known to have an antiferromagnetic transition at $\sim$ 15K \cite{Busch1965}. We show NdSb does possess XMR of order $10^4\%$ which is comparable to other topological semi-metals. We also show that this type-I fcc antiferromagnet has a complex $H$-$T$ phase diagram at high field, with multiple first order transitions.

\section{Experimental and Computational Techniques}
Single crystals of NdSb were grown out of Sn flux \cite{Canfield1992}. Electrical contacts for resistivity measurements were made on freshly cleaved samples using \SI{25}{\micro\metre} platinum wires spot-welded to the (001) face with current applied in the (100) direction. Specific heat measurements were performed using the time-relaxation method. In all the measurements discussed here, field was applied in the (001) direction. The temperature was controlled within a Quantum Design physical properties measurement system.

First principles density functional theory calculations were performed to compute the electronic structure using the WIEN2K code \cite{BlahaP.SchwarzK.MadsenG.K.H.KvasnickaD.Luitz2001}. The PBE functional \cite{Perdew1996} under the generalized gradient approximation was employed, and spin orbit coupling was included via a second order variational scheme.  LaSb and NdSb were computed using the rock salt structure with lattice parameters of $a$ =  6.499 $\rm{\AA}$ and 6.319 $\rm{\AA}$, respectively \cite{Samsonov1974, Abdusalyamova1990}. As the $f$-moments of Nd are strongly localized we treated three $f$-electrons on Nd as core electrons.

\section{Results and Discussion}
\subsection{Magnetoresistance}
\begin{figure}[h]
 \includegraphics[width=0.99\columnwidth]{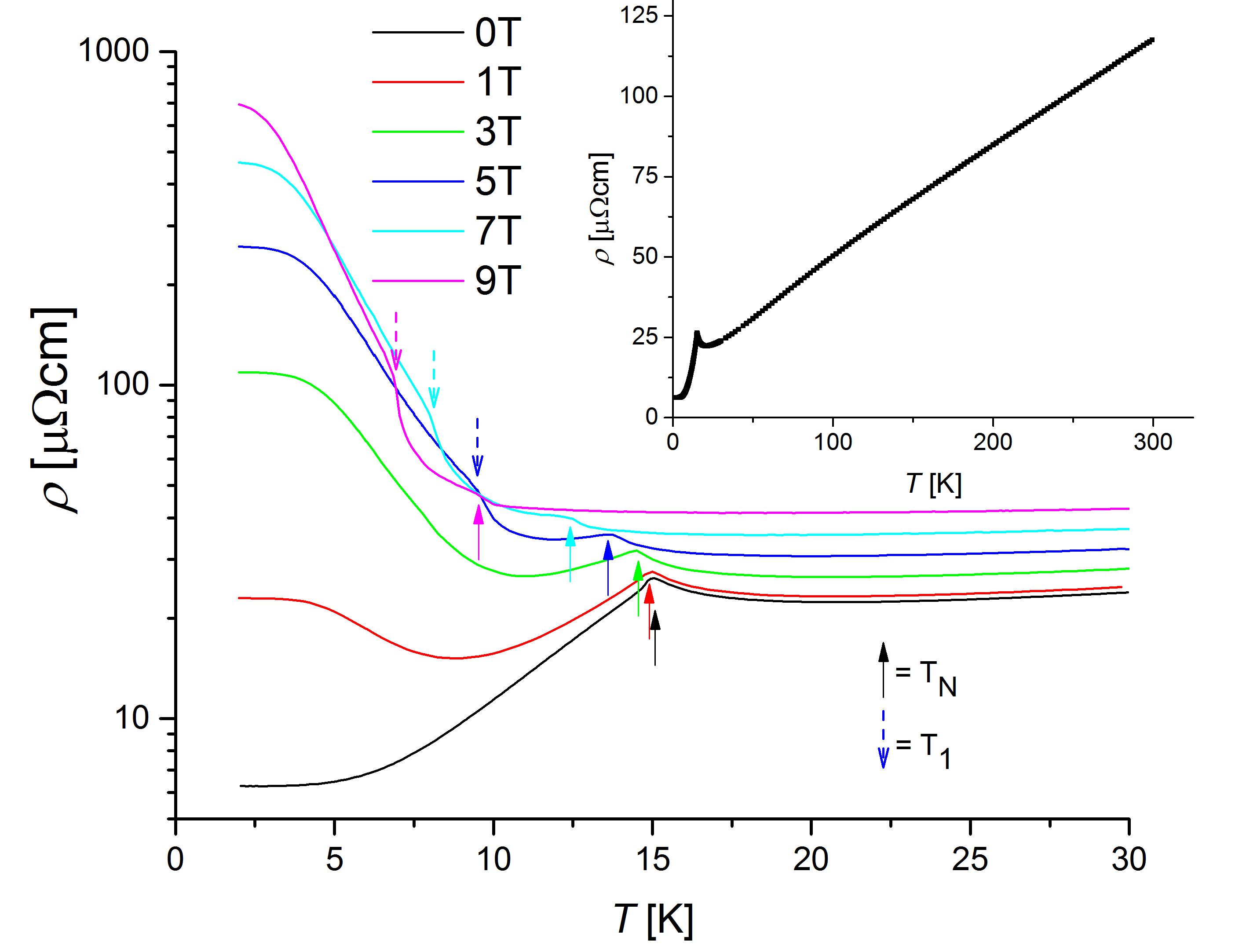}%
 \caption{(Color online) Resistivity of NdSb as a function temperature at various applied magnetic fields. Solid arrows indicate the N$\acute{e}$el temperature $T_N$, dashed arrows show $T_1$. Field was applied in (001) direction perpendicular to the current, which was applied in the (100) direction. Inset shows zero field resistivity up to 300\,K. \label{RvsT}}
 \end{figure}

Figure \ref{RvsT} shows the resistivity $\rho$ of NdSb as a function of temperature $T$ at several different applied magnetic fields. The inset shows $\rho(T)$ in zero field up to room temperature, which demonstrates a clear peak in the resistivity at the previously reported N$\acute{e}$el temperature $T_N = 15$\,K \cite{Busch1965}. It should be noted that the sensitivity of the resistivity to the magnetic order demonstrates that spin disorder scattering provides a significant contribution to the total resistivity close to $T_N$. The absence of a resistive anomaly at 3.7 K (the superconducting transition temperature of Sn) rules out the possible presence of Sn inclusions in our sample.

In an applied magnetic field $T_N$ is suppressed to lower temperature. Above 3T an additional kink appears in the resistivity significantly below $T_N$ at a temperature we define as $T_1$. This kink in the resistivity shows that even at finite fields, and temperatures well below $T_N$, there is still a large contribution to the resistivity from spin-scattering.  In field, at low temperatures the resistivity increases but saturates below 5\,K. This behavior is similar to other XMR materials, such as LaBi and LaSb \cite{Sun2016,Tafti2016,Tafti2015}.

There is yet to be a conclusive explanation of the resistivity upturn in field, and the low temperature saturation in these XMR materials. Evidence has been presented that the large magnetoresistance is the result of the breaking of time reversal symmetry leading to a loss of topological protection against electron back-scattering \cite{Liang2014,Tafti2015}. 
In such a scenario, the low temperature saturation is then caused by the electrical shorting effect of conductive topological surface states. However, it has also been argued that in LaBi the large magnetoresistance and resistivity saturation can be explained by a two band model, where each band has a high mobility, and the density of electrons and holes is comparable \cite{Sun2016,Ali2014}. Such an explanation would not be related to the non-trivial topology of the band structure.

\begin{figure}[h]
 \includegraphics[width=0.99\columnwidth]{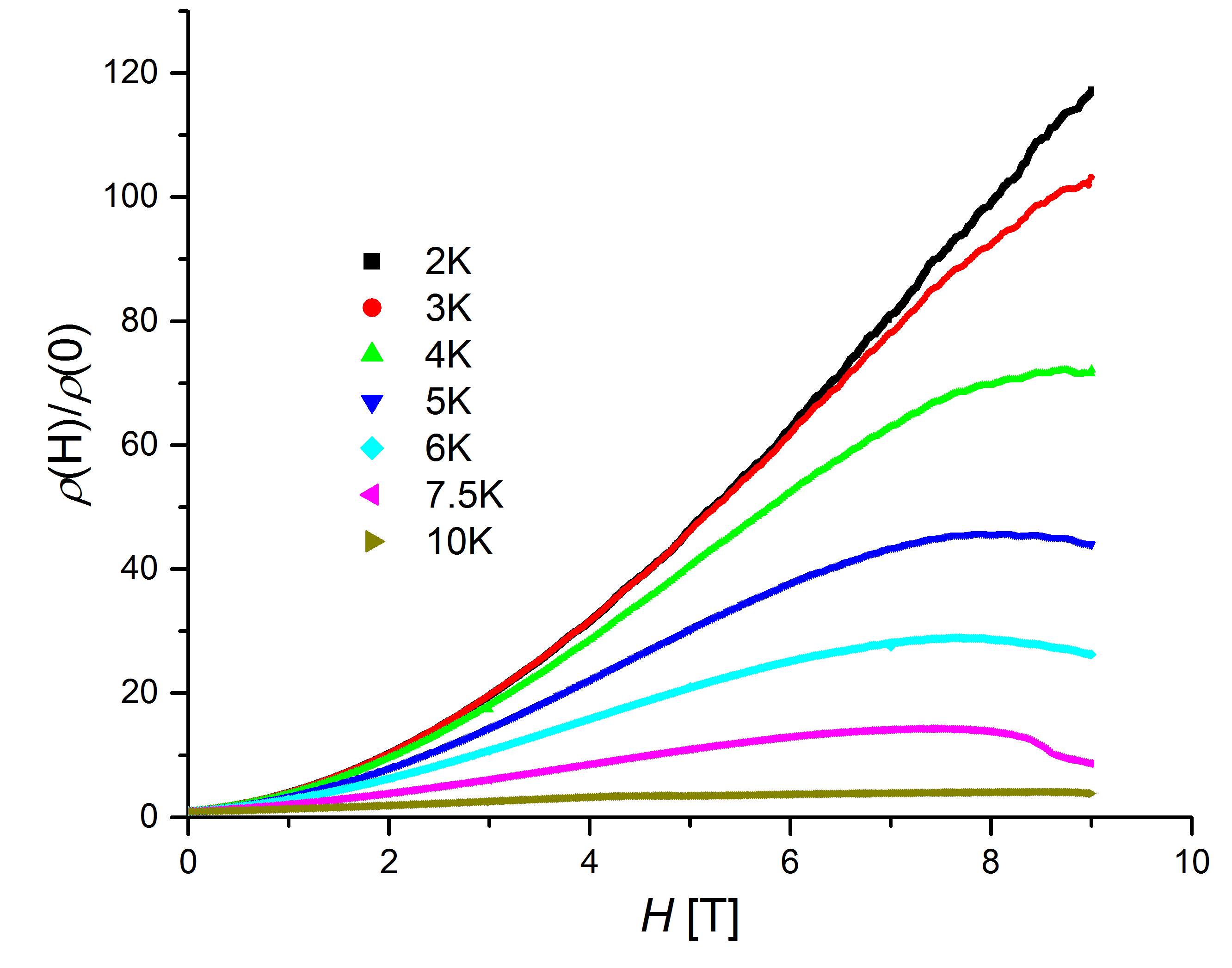}%
 \caption{(Color online) Magnetoresistance of NdSb as a function of applied magnetic field $H$ at various temperatures. Field was applied in (001) direction perpendicular to the current, which was applied in the (100) direction.\label{MR}}
 \end{figure}

Figure \ref{MR} shows the magnetoresistance as a function of field $H$ at several different temperatures. Note the large magnetoresistance of \SI{1.2E4}\% at 2K and 9T. This is smaller than reported in some other XMR materials but is comparable to TaAs and Cd$_3$As$_2$, when the residual resistivity and the residual resistivity ratio is comparable to our sample \cite{Yang2015,Liang2014}. We therefore expect even larger magnetoresistance in cleaner rare-earth monopnictide samples. Above 7.5K there is a clear kink in $\rho(H)$ at high fields. This corresponds to the kink in $\rho(T)$ observed at $T_1$.

\begin{figure}[h]
 \includegraphics[width=0.99\columnwidth]{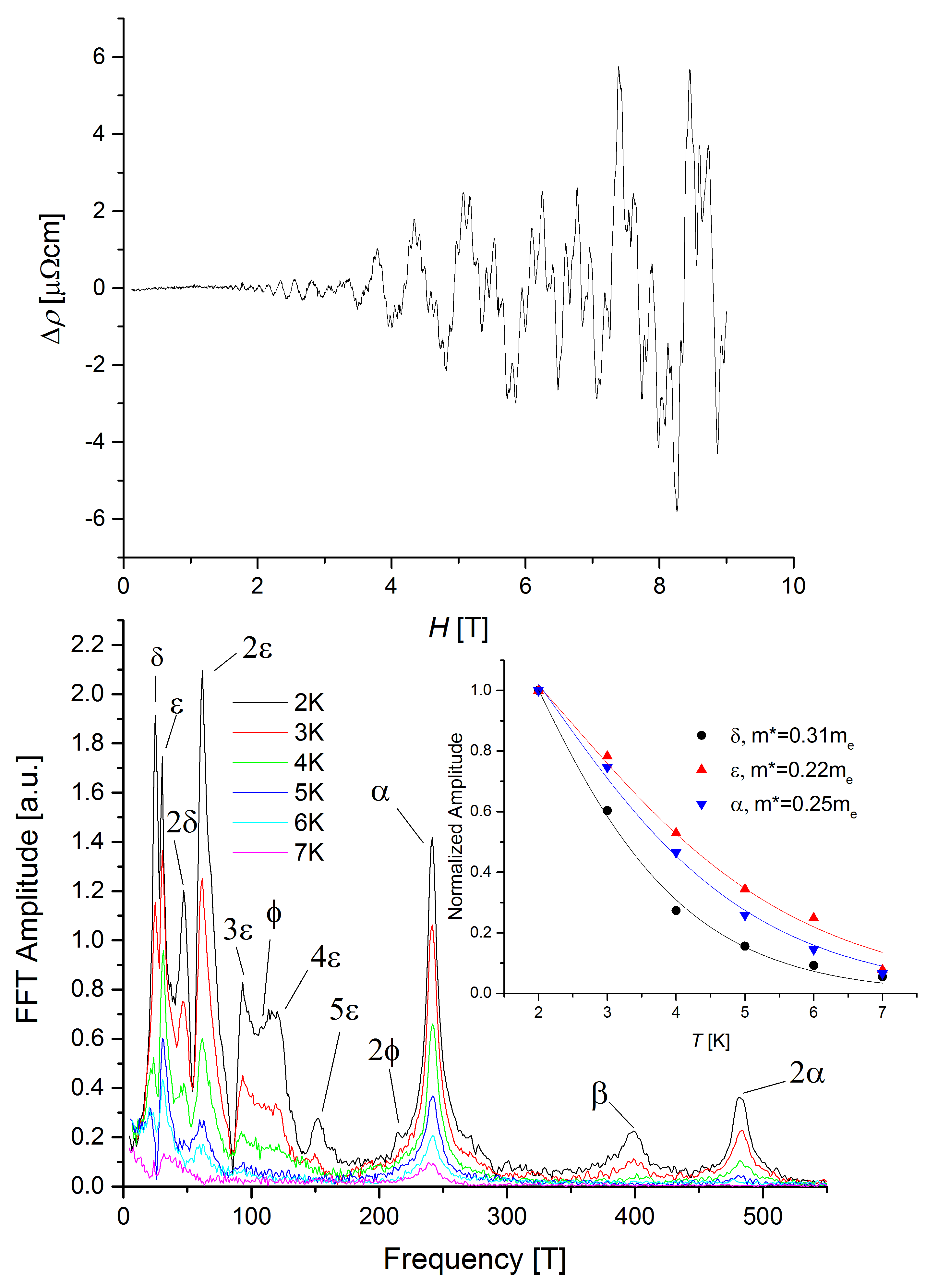}%
 \caption{(Color online) (a) Oscillatory part of the resistivity $\Delta\rho$ as a function of magnetic field $H$ measured at 2\,K. (b) Fast Fourier transform of measured quantum oscillations for various temperatures. The fundamental and harmonic peaks have been identified. The points in the inset show the measured FFT amplitude, normalized at 2K, as a function of temperature for 3 different orbits. The lines are fits to the Lifshitz-Kosevich formula and the quoted masses are calculated from the parameters of this fit. \label{QO}}
 \end{figure}

Quantum oscillations can be observed in the MR at low temperatures and high fields. The oscillations were isolated from the non-oscillatory background by subtraction of a 7th order polynomial fitted to the data. The oscillating contribution to the resistivity $\Delta \rho$ as a function of field is shown at 2K in Fig. \ref{QO}a.

Figure \ref{QO}b shows the amplitude of the fast Fourier transform (FFT) of $\Delta \rho$ at several temperatures, as a function of oscillation frequency. The peaks labeled $\alpha$ and $\beta$ observed at 242T and 397T, respectively, are likely to correspond to peaks measured at 212T and 433T in LaSb \cite{Tafti2015}. We have also labeled $\delta$, $\epsilon$, and $\phi$ peaks at lower frequencies as listed in Table \ref{orbits}. The suppression of the FFT amplitude of the $\alpha$, $\delta$ and $\epsilon$ peaks is clearly resolved with increasing temperature and is plotted in the inset to Figure \ref{QO}b. These data have been fitted to the Lifshitz-Kosevich formula, and an effective mass extracted. We find $m^*_{\alpha}=0.25m_e$, $m^*_{\delta}=0.31m_e$ and $m^*_{\epsilon}=0.22m_e$. The mass extracted from the $\alpha$ peak is in reasonably close agreement with that reported for LaSb ($0.22m_e$) \cite{Tafti2016}, and LaBi($0.22-0.35m_e$) \cite{Tafti2016,Sun2016}.

\begin{figure}[h]
 \includegraphics[width=0.99\columnwidth]{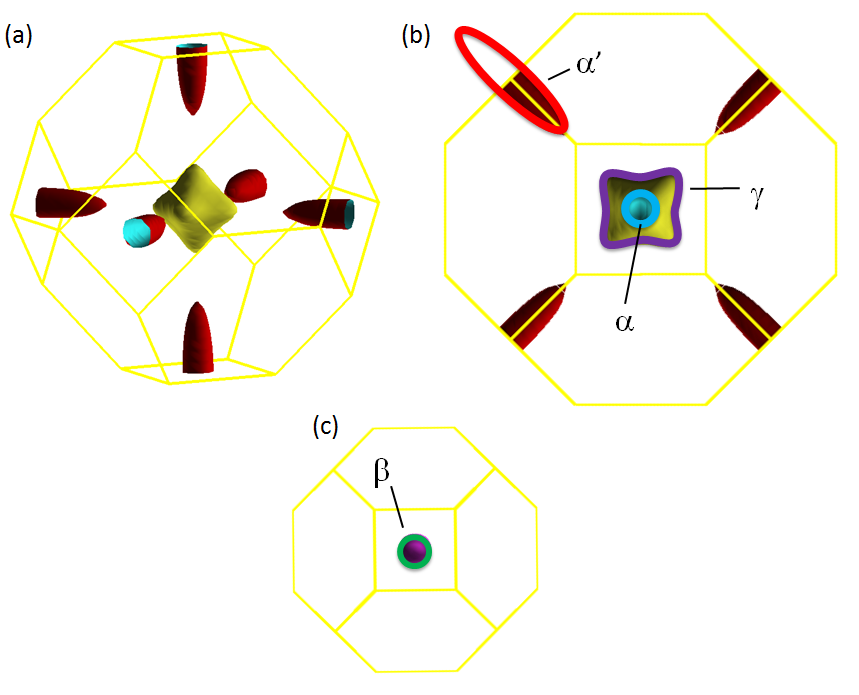}%
 \caption{(Color online) (a) Calculated Fermi surface of NdSb. (b) Projection of FS perpendicular to (001) direction. (c) Same projection as (b), now only showing the central FS pocket. Solid colored lines in (b) and (c) show orbits of different quantum oscillation frequencies. Blue $= \alpha$ orbit, red $= \alpha^{\prime}$ orbit, green $= \beta$ orbit, purple $= \gamma$ orbit.\label{FS}}
 \end{figure}

In order to investigate the origin of the observed oscillation frequencies we have calculated the Fermi surface (FS) of NdSb and LaSb using density functional theory. The calculated FS of NdSb in the paramagnetic state is shown in Fig. \ref{FS}.
The projection perpendicular to the (001) direction is shown in Fig. \ref{FS}b and \ref{FS}c. There are two hole pockets centered at the $\Gamma$ point. The smaller spherical hole-pocket, shown in Fig. \ref{FS}c, is enveloped by the larger yellow octahedral-like pocket in Fig. \ref{FS}a and \ref{FS}b. Finally, an electron pocket forms an ellipsoidal Fermi surface which intersects the Brillouin zone near the ($\pi$,0,0) point and five other high symmetry points related by the cubic symmetry. The electronic structure close to the Fermi energy in NdSb is almost identical to both the calculated and measured Fermi surfaces of LaSb \cite{Kitazawa1983, Hasegawa1985}. For a magnetic field applied along (001) as we have done, four frequencies are expected as labeled in Fig. \ref{FS}b and \ref{FS}c. In LaSb, De Haas-van Alphen (dHvA) measurements have been able to resolve all 4 extremal orbits \cite{Kitazawa1983}, while Shubnikov de Haas (SdH) measurements \cite{Tafti2015} only observe the smaller two frequencies. Hence, we are not surprised that we also only observe the $\alpha$ and $\beta$ orbits in our SdH measurements. We find reasonable agreement between the calculated and observed frequencies for the $\alpha$ and $\beta$ orbits. However, we do note that the agreement is not as good for NdSb as for LaSb. This difference may be related to a small tetragonal distortion, which occurs upon entering the magnetically ordered state \cite{Levy1969}, and/or the influence of the Nd magnetism on the electronic structure.

Note, that without including long range magnetic order there is no way to account for the proliferation of observed frequencies below 200 T. At least 3 unique frequencies labeled $\delta$, $\epsilon$, and $\phi$ can be identified. The additional peaks can be assigned to harmonics of these three frequencies, though they may also contain additional orbits as well. The origin of these additional frequencies not observed in the La-analogs is a natural consequence of the Brillouin zone folding, which occurs due to the long range antiferromagnetic order. Confirmation of this would require data to higher fields in order to separate the frequencies present above and below $T_N(H)$.


\begin{table}[ht]

\centering 
{
\renewcommand{\arraystretch}{1.4}
\resizebox{\columnwidth}{!}{
\begin{tabular}{c|| c| c c c c c c c } 
\hline
&Orbit&$\alpha$&$\alpha^\prime$&$\beta$&$\gamma$&$\delta$&$\epsilon$&$\phi$\\ [0.5ex] 
\hline \hline
LaSb & Exp. (dHvA)\cite{Kitazawa1983} &236&923&480&1189&&&\\ 
& Exp. (SdH)\cite{Tafti2016}& 212 & - & 433  & - &  &  &   \\
& Theory  & 224 & 1008 & 480 & 1332 &  &  & \\
\hline
NdSb & Exp. (SdH) & 242 & - & 397  & - & 24 & 30 & 107  \\ 
& Theory  & 296 & 1380 & 736 & 1728 &  &  & \\ [1ex]
\hline 
\end{tabular}
}
}
\caption{Quantum oscillation frequencies of experimental Shubnikov-de Haas (SdH) and De Haas-van Alphen (dHvA) measurements, and theoretical orbits for LaSb and NdSb. Frequencies are shown in units of T. } 
\label{orbits} 
\end{table}

\subsection{Antiferromagnetism}
\begin{figure}[h]
 \includegraphics[width=0.99\columnwidth]{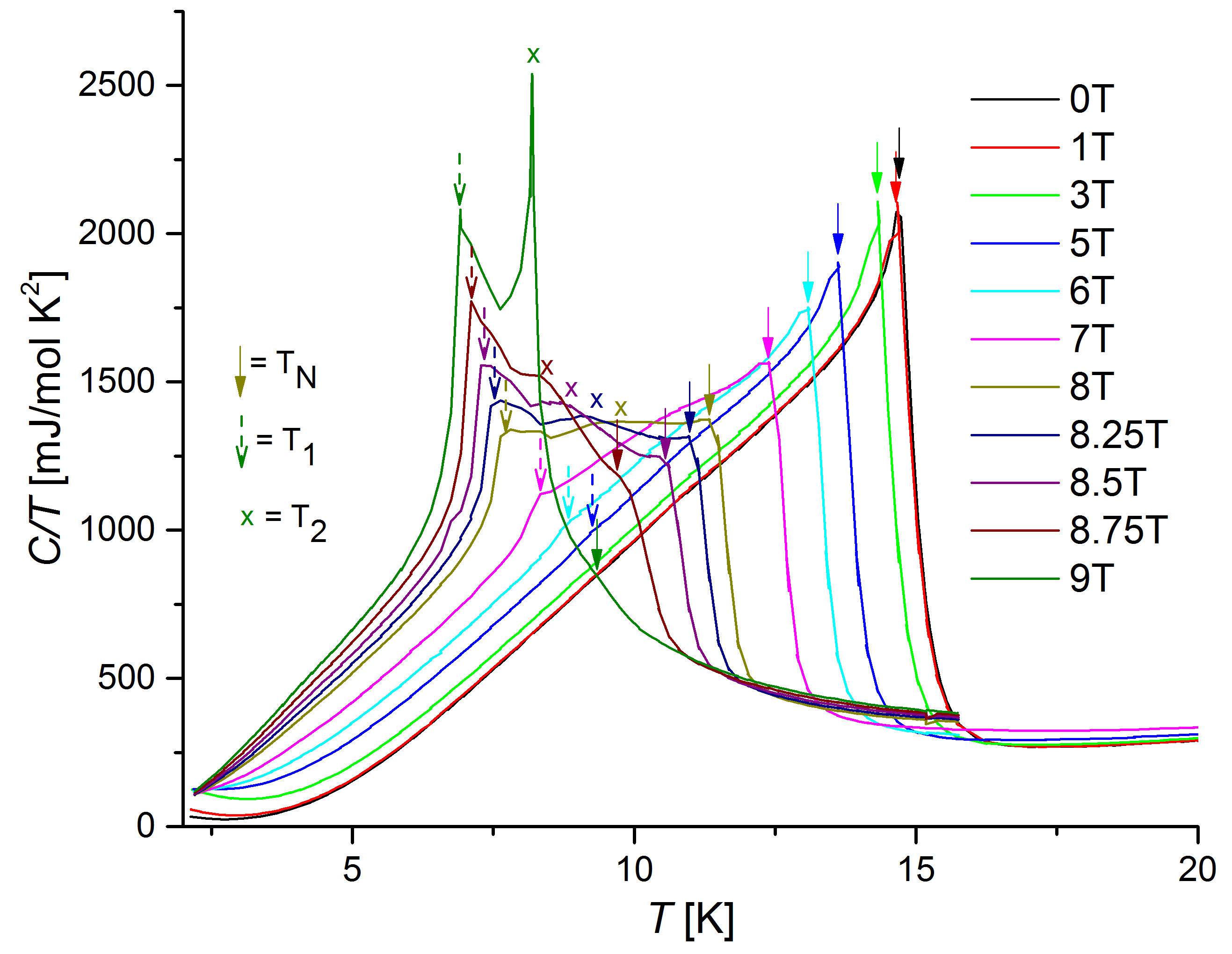}%
 \caption{(Color online) Specific heat $C$ divided by temperature $T$ of NdSb as a function of $T$ at various applied magnetic fields. Solid arrows indicate the N$\acute{e}$el temperature $T_N$, dashed arrows show $T_1$, and crosses indicate $T_2$. Field was applied in (001) direction. \label{CoverT}}
 \end{figure}

NdSb is a type-I fcc antiferromagnet with ferromagnetically aligned planes of Nd moments, antiferromagnetically coupled in the (001) direction. The moments are aligned in the (001) direction \cite{Schobinger-Papamantellos1973}. The antiferromagnetic transition can clearly be seen as a peak in the specific heat divided by temperature $C/T$ as a function of temperature, shown in Fig. \ref{CoverT}.
The magnetic entropy up to $T_N$ estimated from this data is $0.85R\ln(4)$, where R is the universal gas constant. This is in good agreement with that expected for this system, where the f-electrons of Nd are localized, and the groundstate of the system is a quartet \cite{Schobinger-Papamantellos1973, Furrer1976}. This model of a localized moment is also consistent with the observation that the ordered moment in NdSb is very close to the free ion value \cite{Schobinger-Papamantellos1973}. These observations justify the treatment of the Nd moments as localized in our FS calculations.

In a magnetic field applied in the (001) direction $T_N$ decreases with increasing field, as expected. In addition, at fields above 5T there is a second feature in $C/T$ below $T_N$ marked by a dashed arrow in Fig. \ref{CoverT}. For example, at 7T there is a kink in $C/T$ at around 8.3\,K. This coincides with the kink seen in the resistivity shown in Fig. \ref{RvsT} at $T_1$. As the field is increased further to 9T the kink in $C/T$ becomes a peak at 7K and the transition becomes strongly first order. The antiferromagnetic transition becomes less pronounced at high fields, and is only a very small kink at 9.8\,K at 9\,T. There is also a third feature in $C/T$ at a temperature between $T_N$ and $T_1$ marked by crosses, that we define as $T_2$. This feature becomes very well defined and first order at 9T at 8.2K. No feature of this transition was observed in resistivity measurements.

\begin{figure}[h]
 \includegraphics[width=0.99\columnwidth]{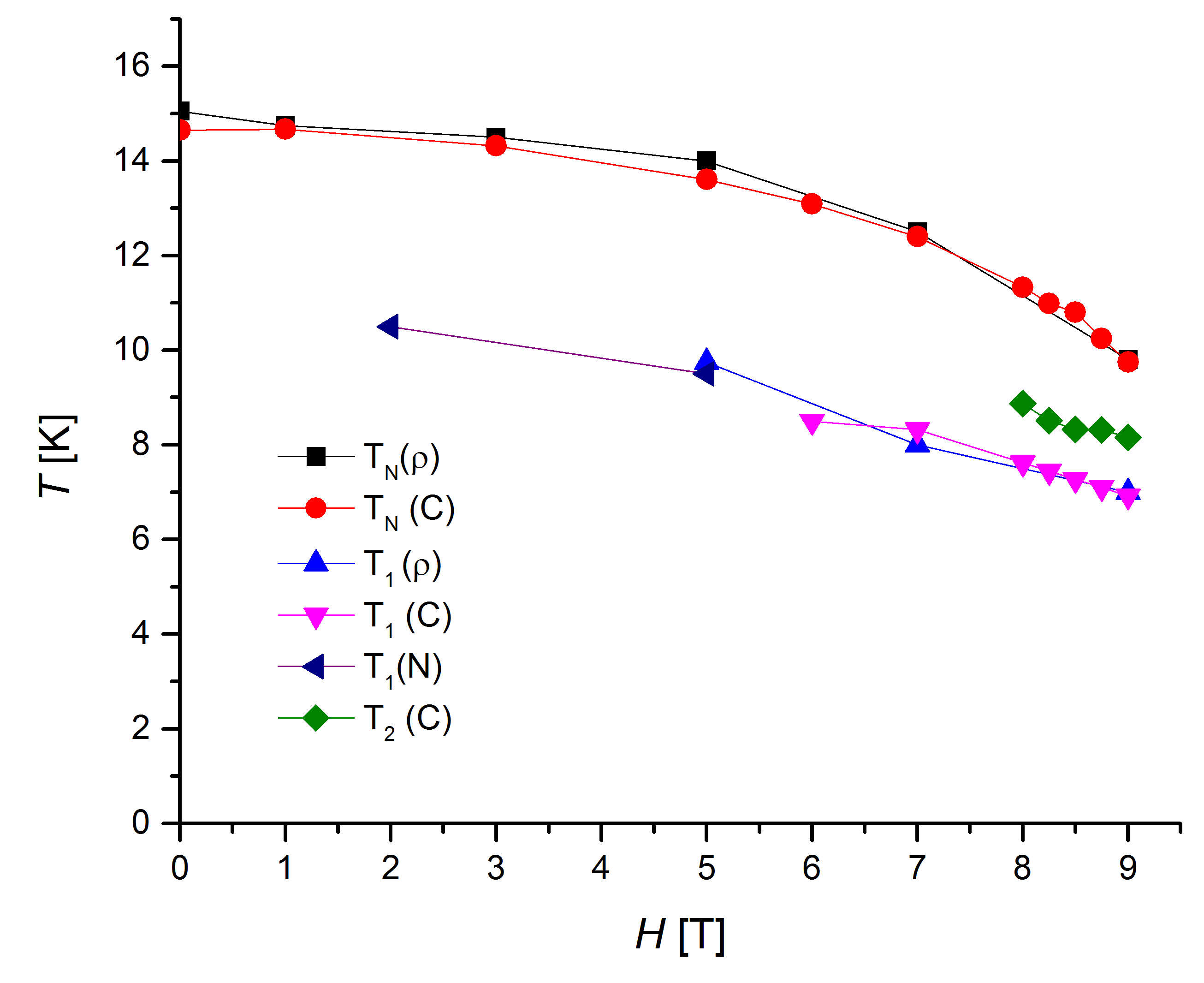}%
 \caption{(Color online) Phase diagram of NdSb as a function of field and temperature, taken from our measurements of resistivity ($\rho$) and  specific heat (C), as well as previous neutron scattering meausurements (N) \cite{Mukamel1985}.  \label{phaseDiag}}
 \end{figure}

There are multiple possible explanations for the complexities of the $H$-$T$ phase diagram in NdSb. From the symmetry of the AFM order, in zero field one would expect equal populations of magnetic domains with the AFM Q-vector along the (100), (010) and (001) directions. However, it was shown from neutron scattering measurements that in NdSb this is not the case. In fact, the population of each domain is different in zero field, and the (001) domain fraction is dominant near $T_N$ when field is applied in the (001) direction \cite{Mukamel1985}. Furthermore, in sufficiently high field as temperature is reduced the population of the (001) domain is suddenly quenched, and the population of the (100) and (001) domains is enhanced. The temperature of this first order transition has been added to our $H$-$T$ phase diagram in Fig. \ref{phaseDiag}. This phase boundary coincides perfectly with the anomalies we observe in resistivity and heat capacity and have labeled as $T_1$. Other than magnetic domain redistribution it was also suggested that this phase transition could involve multi-Q ordering \cite{VogtChapter}. Finally, we note that the first excited crystal field level in NdSb is only 1.2 meV above the ground state energy \cite{Furrer1972}, and hence will play a role in determining the subtleties of the phase diagram.


Figure \ref{phaseDiag} shows the magnetic phase diagram constructed from the resistivity and specific heat measurements of NdSb.


\section{Conclusion}

In conclusion, through measurements of the resistivity and specific heat we have shown that the antiferromagnetic topological semi-metal NdSb has a complex $H$-$T$ phase diagram. Transport measurements also show XMR that is qualitatively and quantitatively consistent with that observed in the non-magnetic lanthanum monopnictide analogues, LaSb and LaBi. This suggests that the large magnetoresistance is not solely the result of breaking TRS on a topological conductive state.  Quantum oscillations measurements demonstrate that the FS of NdSb is similar to LaSb, although with the addition of lower frequencies peaks, which we attribute to zone folding because of the magnetic groundstate. In addition, the transport is sensitive to the spin-disorder scattering which occurs most notably near $T_N$. The persistence of XMR in NdSb, despite the clear influence of magnetism on the resistivity, emphasizes that magnetism can coexist with the high mobilities often found in topological materials.


We thank Yongkang Luo for many helpful discussions. Synthesis and measurements of the crystals were performed under the auspices of the U.S. Department of Energy, Office of Science. Electronic structure calculations were performed with the support of the Los Alamos National Laboratory LDRD program.

\end{document}